\documentclass[conference]{IEEEtran}
\IEEEoverridecommandlockouts

\usepackage{cite}
\usepackage{amsmath,amssymb,amsfonts, mathtools}
\usepackage{algorithmic}
\usepackage{graphicx}
\usepackage{textcomp}
\usepackage{xcolor}
\def\BibTeX{{\rm B\kern-.05em{\sc i\kern-.025em b}\kern-.08em
    T\kern-.1667em\lower.7ex\hbox{E}\kern-.125emX}}
\begin{document}

\title{\huge NANCY: Neural Adaptive Network Coding methodologY for video distribution over wireless networks}


\author{\IEEEauthorblockN{Paresh Saxena, Mandan Naresh, Manik Gupta, Anirudh Achanta}
\IEEEauthorblockA{\textit{Dept. of CSIS, BITS Pilani}\\
Hyderabad, India \\
\{psaxena, p20180420, manik, f21060022\}@hyderabad.bits-pilani.ac.in}
\and
\IEEEauthorblockN{Sastri Kota}
\IEEEauthorblockA{\textit{University Of Oulu} \\
Oulu, Finland \\
sastri.kota@gmail.com }
\and
\IEEEauthorblockN{Smrati Gupta}
\IEEEauthorblockA{\textit{Microsoft Corporation} \\
Redmond, Washington, USA \\
smrati.gupta@microsoft.com}

}

\maketitle

\begin{abstract}
This paper presents NANCY, a system that generates adaptive bit rates (ABR) for video and adaptive network coding rates (ANCR) using reinforcement learning (RL) for video distribution over wireless networks. NANCY trains a neural network model with rewards formulated as quality of experience (QoE) metrics. It performs joint optimization in order to select: (i) adaptive bit rates for future video chunks to counter variations in available bandwidth and (ii) adaptive network coding rates to encode the video chunk slices to counter packet losses in wireless networks. We  present the  design  and  implementation  of  NANCY, and  evaluate  its performance  compared  to  state-of-the-art video rate adaptation algorithms including Pensieve and robustMPC. Our  results show  that  NANCY  provides 29.91\% and 60.34\% higher average QoE than Pensieve and robustMPC, respectively.
\end{abstract}


\section{Introduction}\label{section:introduction}
Internet video services have experienced tremendous growth in last few decades and have become a part of everyday interactions. Based on the technical report from Cisco \cite{cisco20}, the total Internet video is expected to be $79\%$ of all Internet traffic by the end of 2020 with content delivery networks (CDNs) alone to deliver more than 73\% of all Internet video traffic. Moreover, most of the video streaming happens through user devices with wireless connectivity, i.e. 3G/4G cellular or WiFi services. The video demand is expected to be potentially even higher in future 5G networks for more advanced and sophisticated real time video applications like remote medical surgery, augmented reality, mobile broadcasting\cite{Nightingale}. Owing to the huge demand for video delivery services, content providers often struggle to provide high quality video to end-users. Further, the underlying wireless network characteristics including bandwidth variations, latency, packet losses, etc, can highly influence the video quality.  Both these factors, thus necessitate the need for improving the quality of video delivery services over wireless networks.

In order to cope up with varying wireless network conditions, one of the traditional approaches to improve video quality is to make use of adaptive bit rate algorithms \cite{huang12}, \cite{sun16}. The client requests the video from the server of a specific video quality based on the estimated network conditions and past decisions on bit rates. However, these algorithms are usually optimized for specific scenarios where pre-programmed models are used to generate adaptive bit rates to optimize Quality of Service (QoS) or Quality of Experience (QoE) metrics. Recently, the integration of machine learning (ML) techniques for video streaming has been proposed in \cite{Mao19}, \cite{Akhtar18}, \cite{bhattacharyya19}, \cite{saleem19}. However, the prime focus in the earlier techniques has been on application of ML to generate adaptive video quality levels for bandwidth fluctuations that may arise due to the congestion in the network. They do not consider the packet losses due to poor reception and underlying wireless fading conditions such as in remote locations with limited connectivity. Error concealment techniques of video codecs and physical layer adaptive modulation schemes provide protection only against short temporal losses and fails to provide protection against severe losses \cite{Usman15}. 

Network coding (NC) has proven to be a powerful tool to combat packet losses. With respect to packet loss recovery, NC can be seen as a generalization of forward error correction (FEC) codes with an inherent advantage of in-network coding \cite{Yang17}, \cite{Saxena15} for the topologies beyond end-to-end topology. NC allows mixing of packets \cite{Desmond08} so that a fixed amount of additional packets are sent along with the original packets. In the event of packet losses, these additional packets are used to recover the original packets. The network coding rate is selected based on the feedback on packet loss ratio such that the desired residual packet loss ratio is achieved. Usually, the selection of network coding rates depend on the pre-programmed model considering specific assumptions of the underlying systems.

In this paper, a system called Neural Adaptive Network Coding methodologY, NANCY has been proposed that learns and generates adaptive bit rates (ABR) for video and adaptive network coding rates (ANCR) for network coding without considering any pre-programmed model and assumptions of the underlying systems. The main contribution of the current work is to integrate adaptive network coding rates to existing adaptive bit rates in state of the art systems such as Pensieve \cite{Mao19}. NANCY provides an overall comprehensive system to counter both congestion and packet losses arising from the varying network conditions. NANCY uses reinforcement learning (RL) \cite{sutton11} and is trained with rewards that are formulated with QoE metrics. The results show that by incorporating network coding to counter packet losses and reducing re-transmissions, NANCY achieves a higher and more stable video bit rate. Specifically, NANCY  provides an overall 29.91\% and 60.34\% higher average QoE than Pensieve and robustMPC \cite{Yin15} respectively. 

The rest of the paper is structured as follows. Section \ref{section:relatedwork} highlights the related work on use of NC for video delivery systems as well as use of ML approaches for NC. Section \ref{section:nancydesign} presents details about NANCY design and system architecture along with background on NC and RL approaches applied in the current work. Section \ref{section:measurementsetup} explains in detail the measurement setup along with the performance metrics used for comparison with existing algorithms. Section \ref{section:results} presents the comparison results and Section \ref{section:conclusion} draws conclusions on the paper along with possible future research directions.

\section{Related Work}\label{section:relatedwork}

There are several proposals leveraging the benefits of NC together with video delivery including \cite{nguyen10}, \cite{esmaeilzadeh16}, \cite{pimentel13} and \cite{vukobratovic2018}. Hierarchical Network Coding (HNC) technique is proposed in \cite{nguyen10} for using NC with scalable video bit stream in CDNs. Further, \cite{esmaeilzadeh16} and \cite{pimentel13} investigate the use NC for the unequal error protection (UEP) for layered video delivery. Recently, \cite{vukobratovic2018} proposed the integration of network coding with 5G for mobile video delivery. However, all the existing works focus on the pre-programmed network coding models. They do not explore the novel ways for incorporating learning techniques to adapt network coding parameters. 

Some of the recent proposals on using machine learning along with network coding include \cite{Matsumine19}, \cite{Almanza19}, \cite{Jabbarihagh07} and \cite{nguyen17}. The application of machine learning on physical layer network coding on signals is proposed in \cite{Matsumine19} with the aim on achieving higher throughput, while \cite{Almanza19} investigates the use of ML on designing the coordinator server to synchronize all nodes, and to assign their roles during a NC operation. Further, \cite{Jabbarihagh07} proposed the role of ML for construction of network codes and \cite{nguyen17} investigates the integration of ML with NC for wireless broadcast. 

The proposed work is different from the above as it focuses on the integration of reinforcement learning with the application layer level NC for recovering from packet losses by dynamically adjusting NC rates. The proposed approach utilizes ML techniques for the overall video delivery system optimization by integration of video and network coding components. Using this approach, we jointly predict the parameters for both video and NC by considering different input parameters including feedback on throughput prediction, packet losses, etc. Additionally, the results are based on the comparison of QoE metrics, thus providing the analysis of a complete end-to-end system.
\section{Overall NANCY Design and System Architecture}\label{section:nancydesign}

\begin{figure*}[h]
  \centering
  \includegraphics[width=170mm,scale=2]{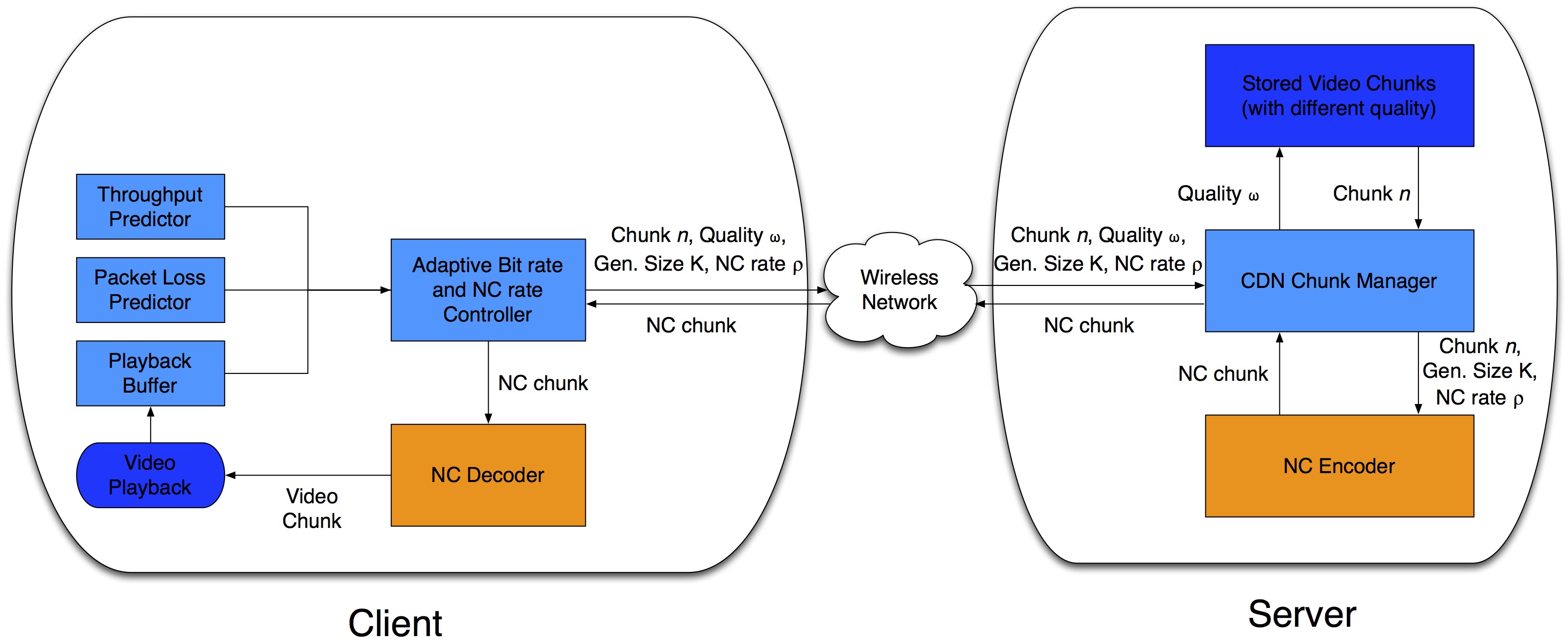}
  \caption{Overall NANCY System Architecture}
  \label{fig:systemmodel}
\end{figure*}

NANCY is designed predominantly for video content delivery and the system architecture is based on the adaptive video streaming design such as Dynamic adaptive streaming over HTTP (DASH) \cite{Stockhammer11}. The proposed system design is shown in Figure~\ref{fig:systemmodel} where videos are stored as multiple chunks at server encoded at different video quality levels. The client requests for a specific chunk with desired video quality from server. The client's decision on chunk request is based on the predicted throughput and playback buffer size. NANCY incorporates the NC functional blocks at both server and client sides for improved system performance. The server has a Content Delivery Network (CDN) Chunk Manager that passes the source chunk to NC Encoder where NC Encoder generates coded chunk. The coded chunk is then sent to the client. The coded chunk is decoded by NC Decoder at the client side and the source chunk is thus delivered to the video player. 
The estimation of video parameters and network coding parameters is done by RL agent referred as Adaptive bit rate and NC rate Controller in Figure~\ref{fig:systemmodel}. The request based on these parameters is sent to the server. Specifically, the request consists of chunk index $n$, generation size $K$, network coding rate $\rho$ and video quality $\omega$. After receiving the request, the server sends a network coded chunk $n$ with desired video quality $\omega$ to the client. The process of network coding and role of network coding rate $\rho$ is explained in the following subsection.  

\subsection{Overview of Network Coding}

An overview of the NANCY's network coding functions is described as follows. Let us assume that the video file is segmented into $n$ chunks and denote the $j$-th~chunk as~$C_{j}$, with its size as~$n_{c}^{j}$ (in bytes) where $j=1,2,\dots,n$. Each chunk is further encoded into slices. The $k$-th~slice is denoted as $S_{k}$ and its size as $n_{s}$ (in bytes) where $k=1,2,\dots,m$ with $m=\lceil{\frac{n_{c}^{j}}{n_{s}}}\rceil$ as the number of slices per chunk. Accordingly, $C_{j} = \begin{bmatrix} S_{1}& S_{2}& .& .& S_{m}\end{bmatrix}$. This is called as 'slice NC generation' a set of $K$~slices, denoted as~$X \in \mathbb{F}_{q}^{n_{s} \times K}$. Here per-slice NC generation block coding with block length~$N$ is assumed. Lets denote $\rho = \frac{K}{N}$ as slice-level coding rate. The encoding process is linear such that the coded packets are given by $Y=XG$ where $Y$ $\in$ $\mathbb{F}_{q}^{n_{s} \times N}$ is a set of $N$ coded slices and $G \in \mathbb{F}_{q}^{K \times N}$ is the generation matrix for network coding. Specifically, a systematic coding approach is followed where $N$ coded slices also consist of $K$ original slices. Hence, the generator matrix $G$ results in $G=[I_{K}  C]$ where $I_{K}$ is the identity matrix of size $K$ and $C \in \mathbb{F}_{q}^{K \times N-K}$ is the matrix with $K(N-K)$ coefficients. There are several deterministic and random ways of selecting these coefficients \cite{Desmond08}, \cite{Yang17}, \cite{Saxena15}, etc. In the current work, the coefficients can be randomly selected from the finite field $\mathbb{F}_{q}$. However, since network coding functions are separated from video streaming, the design provides the flexibility of choosing deterministic coefficients as well.  


\subsection{Learning Algorithms for ABR and ANCR}

In the current work, RL has been used to train the agent and select video bit rate and network coding rate. RL is modeled as a Markov decision process with agents states and actions. Lets consider a discrete system where time $t$ is indexed by $t\in\{1,2,...\}$. At each time step $t$, the agent observes some state $s_{t}$ and chooses an action $a_{t}$. The agent moves to state $s_{t+1}$ and receives reward $r_{t}$. The agent selects action based on a policy, $\pi: \pi_{\theta}(s_{t},a_{t}) \rightarrow [0,1]$ where $\pi_{\theta}(s_{t},a_{t})$ is the probability that action $a_{t}$ is taken in state $s_{t}$ and $\theta$ are the policy parameters upon which the actions are based. The goal of RL agent is to collect as much reward as possible and to find the policy $\pi^{*}$ that maximizes the reward. The optimal policy is given by,
\begin{equation}
    \pi^{*} = argmax_{\pi}\mathop{\mathbb{E}}[\sum_{t=0}^{\infty}{\gamma^{t}}{r_{t}}|s_{0}, a_{t}\sim \pi(.|s_{t})]
   \label{eq:optimalpolicy}
\end{equation}
The overall reward is defined by $\mathop{\mathbb{E}}[\sum_{t=0}^{\infty}{\gamma^{t}}{r_{t}}$], where $\gamma \in$ (0,1] is a factor discounting future rewards.

\begin{figure*}[h]
  \centering
  \includegraphics[width=\linewidth]{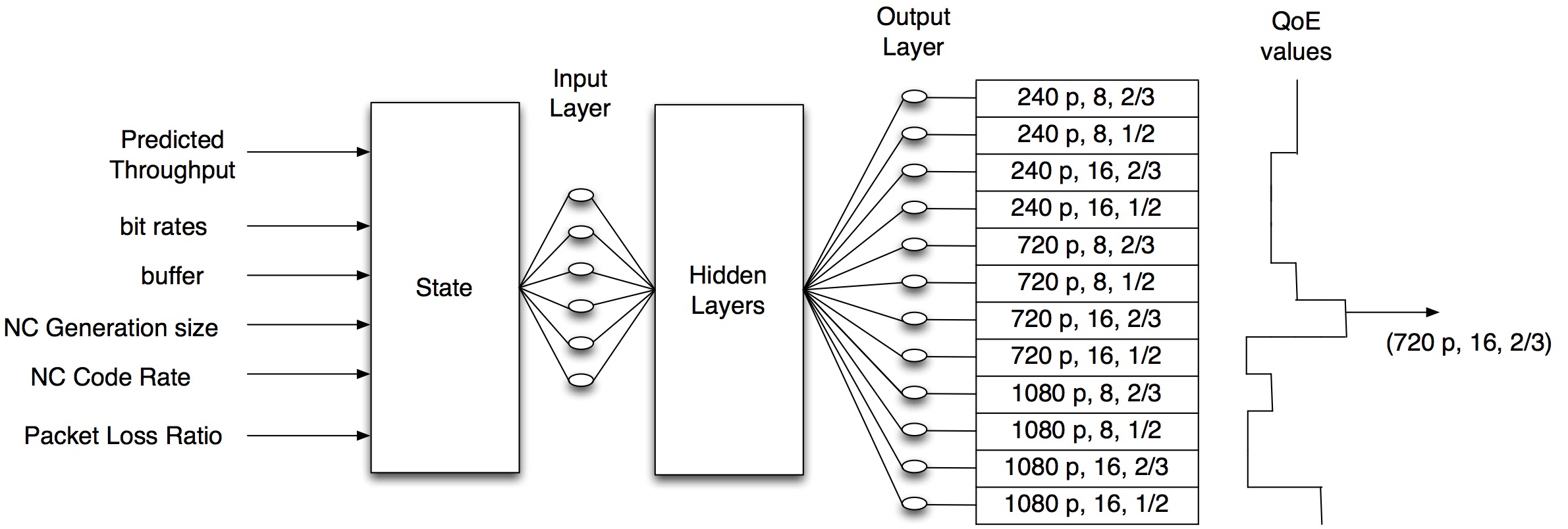}
    \caption{Reinforcement learning to generate adaptive bit rates and adaptive network coding rates}
  \label{fig:adaptivenc}
\end{figure*}

The proposed RL based ABR and ANCR is shown in Figure~\ref{fig:adaptivenc}. By training a neural network, the agent takes an action $a_{t} \in \mathcal{A}$ at every time step $t$ in order to maximize the overall reward. $\mathcal{A}$ consists of different values for $\omega$, $K$, $\rho$, i.e., the code rate $\rho$, generation size $K$ and the video quality $\omega$, as shown in Figure~\ref{fig:adaptivenc}. The agent observes the inputs including predicted throughput, past bit rate decisions, buffer occupancy, past decisions on NC generation size, NC code rate and packet loss ratio. The agent uses the reward information to train and further improve the model. In the model, the reward is the QoE metrics that describes the overall user's satisfaction on the video delivery. In the next section,  QoE metrics have been defined and its has been described how the performance analysis and results are based on those metrics.

Following the training model in \cite{Mao19}, NANCY is trained using an actor-critic method A3C \cite{Mnih16}. A3C, a policy gradient method, takes advantage of value-based and policy-based RL methods, where actor computes an action based on a state and critic produces expected total reward. The critic network helps the actor network to make ABR and ANCR decisions. Specifically, the update of policy parameters $\theta$ follows the policy gradient with entropy regularization \cite{Mnih16} as,

\begin{equation}
  \theta \leftarrow \theta+\alpha\sum_{t} \triangledown_{\theta} \log\pi_{\theta}(s_{t},a_{t})A(s_{t},a_{t})+\beta \triangledown_{\theta}H(\pi_{\theta}(.|s_{t}))
  \label{eq:Thetaupdate}
\end{equation}
where,
\begin{itemize}
\item $\alpha$ is the learning rate.
\item $A(s_{t},a_{t})$ is the advantage function. It is a measure of how much a certain action a good or bad decision given a certain state. 
\item $\beta$ is to enhance exploration. It is set to a large value initially and decreases while the rewards improves.
\item $\triangledown_{\theta} \log\pi_{\theta}(s_{t},a_{t})$ specifies how to change the policy parameters in order to increase $\pi_{\theta}(s_{t},a_{t})$.
\item H(.) is the entropy of the policy to push $\theta$ in the direction of higher entropy. The higher the entropy, the more random the actions an agent takes.
\end{itemize}
The derivation and further details on Equation (\ref{eq:Thetaupdate}) can be found in \cite{Mnih16}.

\section{Experimental Setup and Performance Metrics}\label{section:measurementsetup}
In order to better understand the impact of different network configurations on NANCY, we now describe our experimental setup that has been used for different measurements together with the performance metrics employed.
\subsection{Experimental Setup}

The experimental setup proposed in \cite{Mao19} has been adapted, wherein the ABR server has been implemented using Python environment. The client queries the server to get the bit-rate for the next chunk of the video. The request query consists of observations on the throughput, playback buffer occupancy, packet losses and other video properties. Based on the observations, the trained model output the bit-rate for the next video chunk. The MahiMahi \cite{Ravi15} framework has been used to emulate network conditions. It is used to record and play web traffic under emulated network conditions. Two different data sets have been used for measurement purposes: broadband dataset provided by FCC and mobile dataset collected in Norway \cite{Riiser13}. These traces are reformatted to be compatible with the MahiMahi framework. Further, different packet loss conditions have been emulated using LossShell component \cite{Ravi15} of MahiMahi emulator to compare the different rate-adaptation algorithms.

\subsection{Performance Metrics and Comparison with Existing algorithms}

In this paper, NANCY has been compared with other rate-adaptation algorithms using QoE metrics. Specifically, three variants of QoE have been selected which have been used in the previous works as well \cite{Spiteri16}, \cite{Yin15} to compare the different rate-adaptation algorithms. The QoE variants are based on the general QoE metric, which is defined as
\begin{equation}
  QoE = \sum_{n=1}^{N}q(R_{n})-\mu\sum_{n=1}^{N}T_{n}-\sum_{n=1}^{N-1}\left|q(R_{n+1}-q(R_{n}) \right| \label{eq:QoEmetric}
\end{equation}
The three components of the QoE metric are explained as follows. The first term includes $R_{n}$ that represents the bit-rate for chunk $n$ and $q(R_{n})$ represents the quality corresponding to bit-rate $R_{n}$. A higher video quality means a higher overall QoE. The second term represents the penalty due to rebuffering time $T_{n}$ and the final term represents the penalty due to fluctuations in video quality that hinders the overall smoothness. The three variants that depend on the above general QoE metric are specified as follows.
\begin{itemize}
    \item $QoE_{1}$ - linear QoE where $q(R_{n}) = R_{n}$ with $\mu=4.3$
    
    \item $QoE_{2}$ - log based QoE where $q(R_{n}) = log(R/R_{min})$ with $\mu=2.66$
    
    \item $QoE_{3}$ - QoE assigning high quality scores to HD bitrates with $\mu=8$.
    
\end{itemize}

NANCY has been compared with different state-of-the-art rate-adaptation algorithms for all the three QoE variants described above. Specifically, five different rate-adaptation algorithms have been considered for the comparison.

\begin{itemize}
    \item Pensieve \cite{Mao19} is the base case where RL is used as well to train the agent for delivering adaptive bit rates. Pensieve is closest to NANCY with respect to design, however, Pensieve does not include packet loss recovery strategy and hence expected to perform worse in case of packet losses.
    \item robustMPC \cite{Yin15}  optimally combines throughput and buffer occupancy information to produce adaptive bit rates. 
    \item BOLA \cite{Spiteri16} uses Lyapunov optimization techniques to minimize rebuffering for improving video quality. 
    \item Rate-Based (RB) and Buffer-based (BB) algorithms \cite{Bentaleb19} where RB adapts by taking into account only throughput predictions and BB adapts by taking into account only buffer occupancy observations.

\end{itemize}
 
\section{Results}\label{section:results}

In this section, we present the comparison of NANCY with the other state-of-the-art algorithms under packet losses. Both FCC and Norway traces have been used for comparison purposes. In order to take into account the packet loss ratio variations, different scenarios have been considered where packet losses may vary for each trace. Using MahiMahi, a random packet loss ratio for each trace has been selected. The packet loss ratio value is selected from the following set: [0\%, 0.1\%, 0.2\%,...1.8\%, 1.9\%, 2\%] where 0\% and 2\% are also included. 

\begin{figure}[h]
  \centering
  \includegraphics[width=\linewidth]{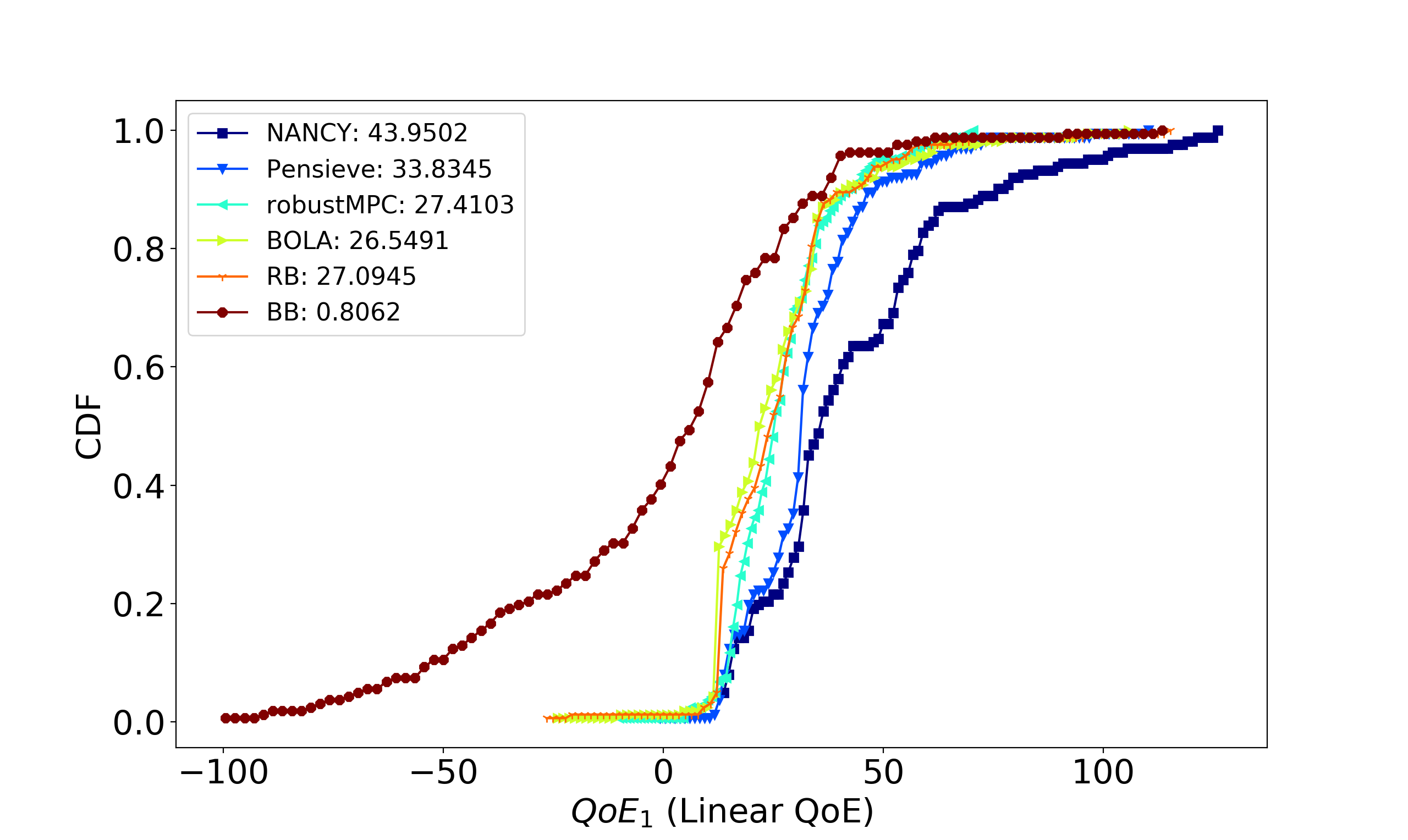}
  \caption{Comparison of NANCY with existing rate-adaptation algorithms for FCC traces with random packet losses within the range of 0\% and 2\%. The legend shows average value of QoE for each algorithm.}

  \label{fig:cdfrandomFCC}
\end{figure}

\begin{figure}[h]
  \centering
  \includegraphics[width=\linewidth]{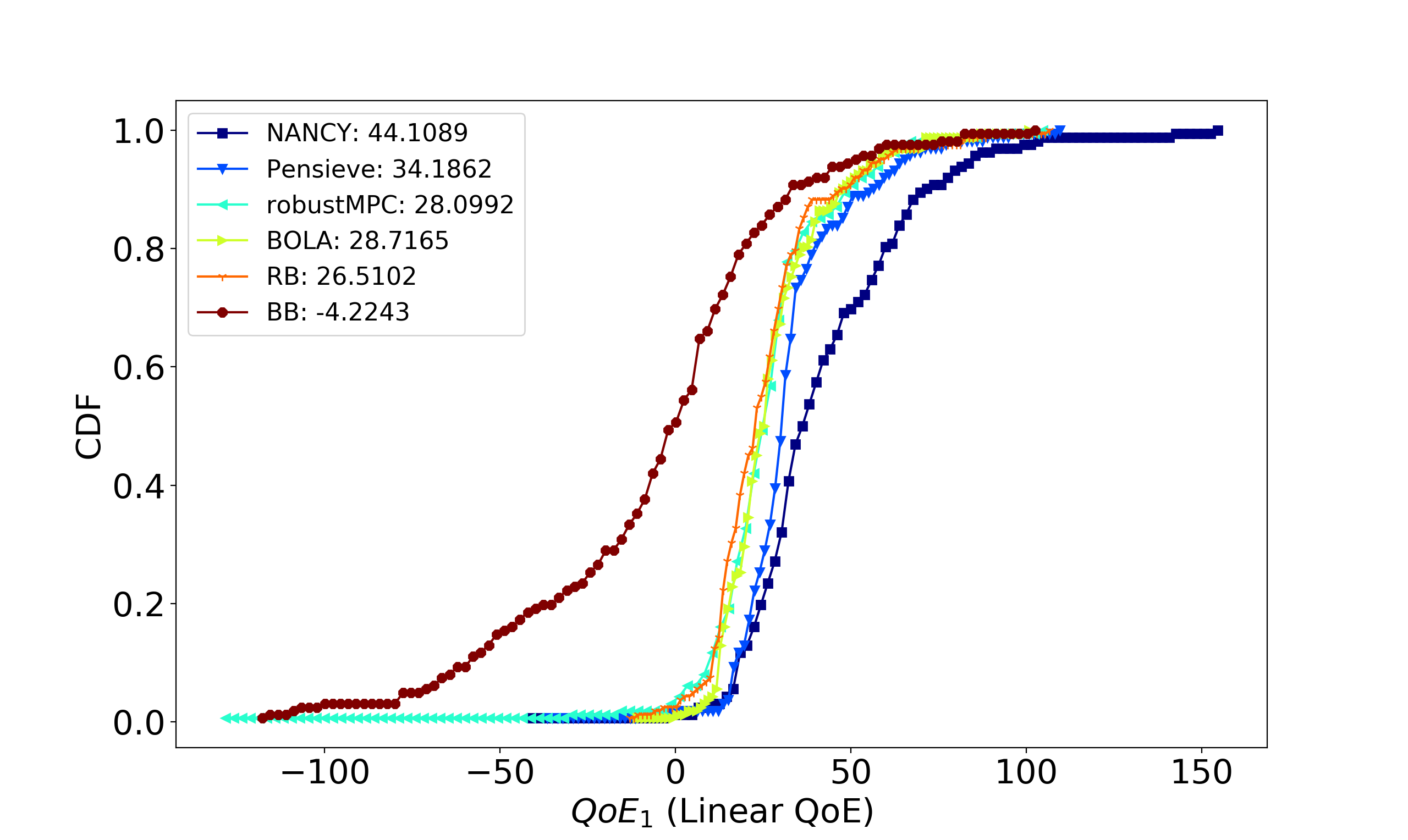}
  \caption{Comparison of NANCY with existing rate-adaptation algorithms for Norway traces with random packet losses within the range of 0\% and 2\%. The legend shows average value of QoE for each algorithm.}

  \label{fig:cdfrandomNorway}
\end{figure}

The results show the comparison of linear QoE ($QoE_{1}$) in Figure~\ref{fig:cdfrandomFCC} for FCC traces and in Figure~\ref{fig:cdfrandomNorway} for Norway traces. The legend also show the average QoE achieved by all the rate-adaptation algorithms. The results show that NANCY achieves a higher average QoE for both FCC and Norway traces. Specifically, for FCC traces, NANCY provides up to 29.91\% higher QoE than Pensieve and almost 60.34\% higher QoE than robustMPC. Similar trends are  observed for Norway traces as well.

\begin{figure}[h]
  \centering
  \includegraphics[width=\linewidth]{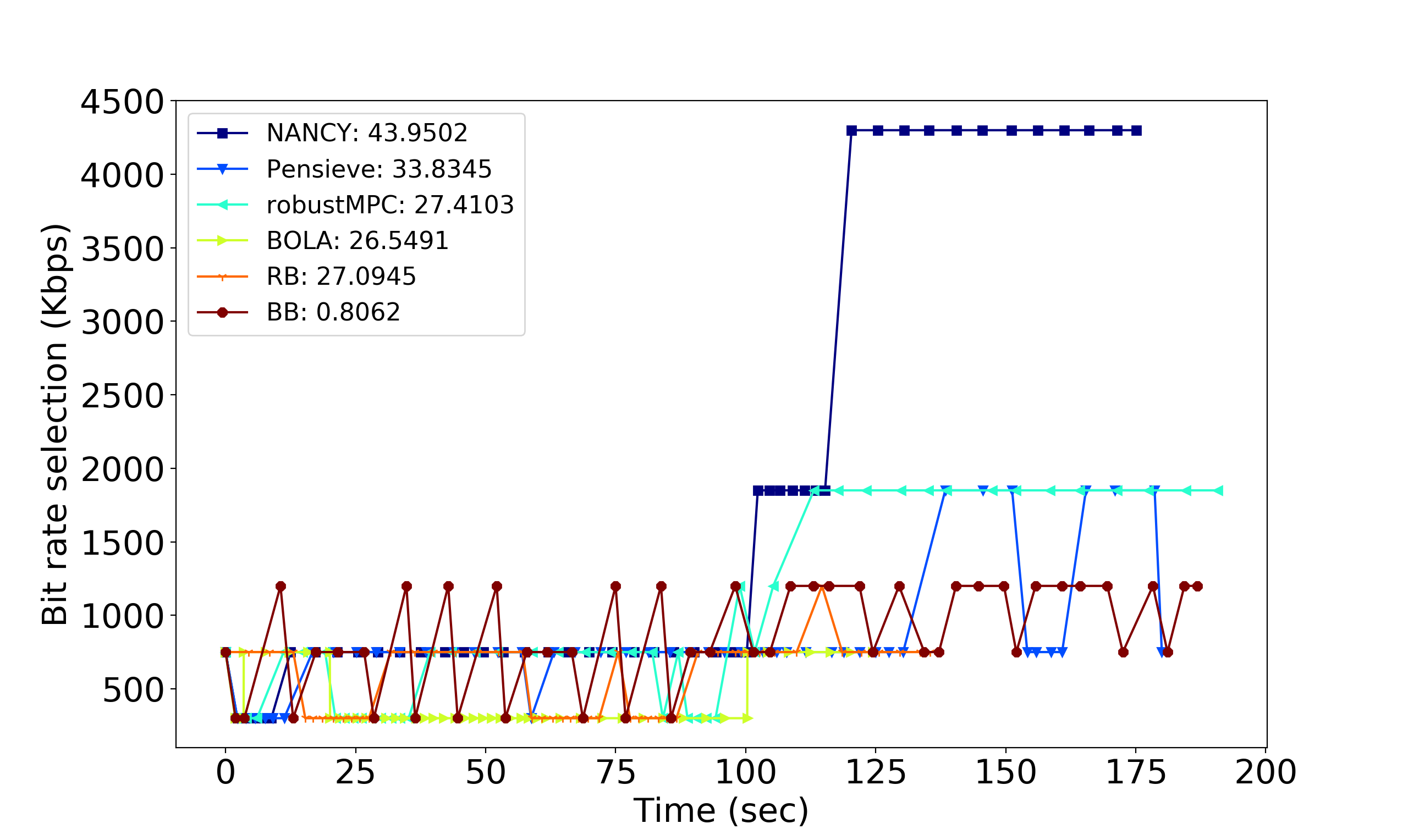}
  \caption{Comparison of bit-rate selection for FCC traces with random packet losses within the range of 0\% and 2\%. The legend shows average value of QoE for each algorithm.}
 
  \label{fig:bitraterandomFCC}
\end{figure}

\begin{figure}[h]
  \centering
  \includegraphics[width=\linewidth]{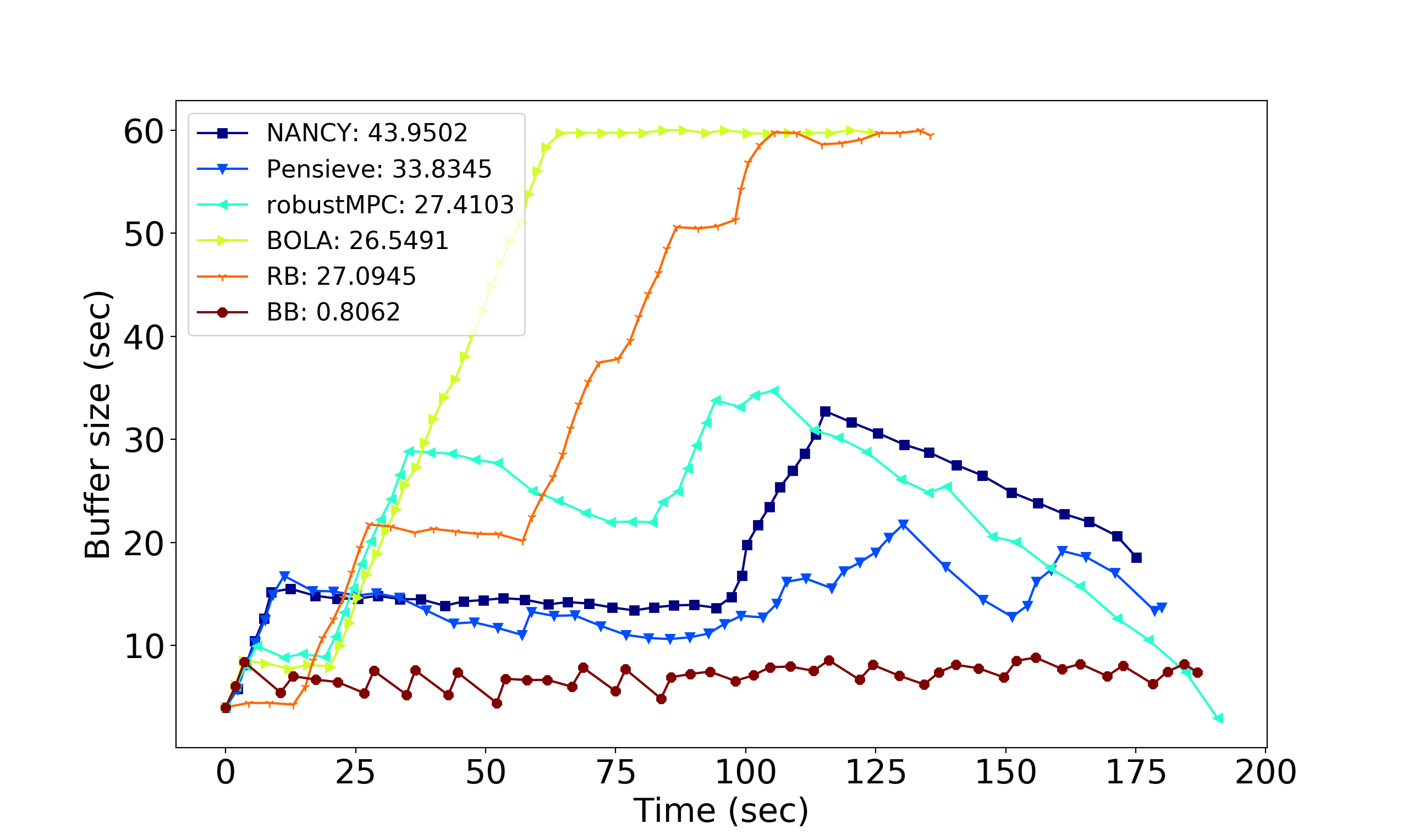}
  \caption{Comparison of buffer size for FCC traces with random packet losses within the range of 0\% and 2\%. The legend shows average value of QoE for each algorithm.}
 
  \label{fig:buffersizerandomFCC}
\end{figure}

\begin{table}
\begin{center}
\begin{tabular}{|c|c|c|}

\hline
\textbf{\textit{ABR algorithms}}& \textbf{\textit{FCC Traces}}& \textbf{\textit{Norway Traces}} \\
\hline
NANCY& 45.62& 41.93 \\
\hline
Pensieve& 35.17& 30.71\\
\hline
robustMPC& 18.70& 27.24\\
\hline
BOLA& 24.50& 24.66\\
\hline
RB& 26.38& 23.02\\
\hline
BB& 1.22& 1.10\\
\hline
\end{tabular}
\caption{Comparison of NANCY with existing rate-adaptation algorithms for FCC and Norway traces using $QoE_{2}$ metric}
\label{tab:QoE2 metric results}
\end{center}
\vspace{-4mm}
\end{table}

\begin{table}
\begin{center}
\begin{tabular}{|c|c|c|}

\hline
\textbf{\textit{ABR algorithms}}& \textbf{\textit{FCC Traces}}& \textbf{\textit{Norway Traces}} \\
\hline
NANCY& 188.69& 210.21 \\
\hline
Pensieve& 123.82& 127.17\\
\hline
robustMPC& 139.15& 142.49\\
\hline
BOLA& 80.26& 80.64\\
\hline
RB& 84.76& 77.10\\
\hline
BB& 31.40& 12.09\\
\hline
\end{tabular}
\caption{Comparison of NANCY with existing rate-adaptation algorithms for FCC and Norway traces using $QoE_{3}$ metric}
\label{tab:QoE3 metric results}
\end{center}
\vspace{-4mm}
\end{table}

In order to better understand the impact of packet losses on QoE, we study the different components of the QoE, specifically, the bit rate selection and the buffer size. Note that both bit rate selection and buffer size impact the QoE value as shown in Equation (\ref{eq:QoEmetric}). Although, a higher bit rate selection increases QoE, but frequent fluctuations and higher buffer size hinder the smoothness of the video impacting the overall QoE. The comparison of bit rate selection and buffer size is shown in Figure~\ref{fig:bitraterandomFCC} and Figure~\ref{fig:buffersizerandomFCC} respectively.   Figure~\ref{fig:bitraterandomFCC} shows that the maximum bit rate achieved by NANCY over the time is more than 4 Mbps whereas Pensieve and robustMPC achieve only around 2 Mbps. The results also show that the bit rate selection in both Pensieve and robustMPC are mostly stable over the time, however, they fail to achieve a higher bit rate as compared to NANCY. Additionally, the bit rate allocation with NANCY is also quite stable for most of the time. The other algorithms including BOLA, RB and BB achieve a smaller bit rate with higher fluctuations as compared to NANCY. 
Similarly, Figure~\ref{fig:buffersizerandomFCC} shows a high buffer size for algorithms including RB and BOLA that results into a higher rebuffering penalty and a smaller QoE. NANCY, Pensieve and robustMPC show similar behavior where NANCY maintains a constant buffer size of a higher duration and hence suffers from smaller rebuffering penalty. 
Hence, it can be inferred from the experiments that overall NANCY achieves a higher bit rate which is stable over a longer period of time and it also encounters a smaller rebuffering penalty. Therefore, it results into an overall higher average QoE as compared to the other rate-adaptation algorithms.

Further, NANCY outperform existing rate-adaptation algorithms for $QoE_{2}$ (Table \ref{tab:QoE2 metric results}) and $QoE_{3}$ (Table \ref{tab:QoE3 metric results}) as well. Our results show that NANCY provides 29.71\% and 36.53\% higher $QoE_{2}$ than Pensieve for FCC and Norway traces, respectively. The gain is even higher for $QoE_{3}$ metric since more weight is given for HD video rates and NANCY is able to achieve higher bit rates. Our results show that NANCY provides up to 52.39\% and 65.29\% higher $QoE_{3}$ than Pensieve for FCC and Norway traces, respectively.  
\section{Conclusion}\label{section:conclusion}
In this paper, the design, development and evaluation of NANCY, a system for generating adaptive video bit rates as well as network coding rates using reinforcement learning has been presented. Working with real traces and emulating packet losses with network emulator, it has been shown that NANCY performs better than the current state-of-the-art video bit-rate algorithms. The results show the benefits of using NANCY for different QoE metrics. The future work includes the performance evaluation of NANCY beyond end-to-end topologies for video multicast and broadcast scenarios. Furthermore, the study will be extended to evaluate NANCY under controlled network conditions to understand the impact of different network parameters including bandwidth, delay, random and bursty packet losses.

\section*{Acknowledgment}

This work has been supported by TCS foundation under the TCS research scholar program and SERB, DST, Government of India's start-up research grant agreement SRG/2019/002027 (MUT-DROCO).

\bibliographystyle{IEEEtran}
\bibliography{sample-base}



\end{document}